\documentclass[12pt]{article}
\usepackage{epsf}
\title{ {\bf Radiative lepton flavor violating decays in the
Randall Sundrum background with localized leptons}}
\author{\vspace{1cm}\\
        {\bf E. O. Iltan}
        \thanks{E-mail address:
        eiltan@newton.physics.metu.edu.tr}
 \\
        Physics Department, Middle East Technical University \\
        Ankara, Turkey\\}

\date{}

\begin{document}
\setlength{\baselineskip}{24pt}
\maketitle
\setlength{\baselineskip}{7mm}
\begin{abstract}
We study  the radiative lepton flavor violating $l_i\rightarrow
l_j\gamma$ decays in the two Higgs doublet model, respecting  the
Randall Sundrum scenario and estimate the contributions of the KK
modes of left (right) handed charged lepton doublets (singlets) on
the branching ratios. We observe that the branching ratios are
sensitive to the contributions of the charged lepton KK modes.
\end{abstract}
\thispagestyle{empty}
\newpage
\setcounter{page}{1}
\section{Introduction}
The flavor violation (FV) is among the most interesting physical
phenomena to search the standard model (SM) and its beyond. The
lepton FV is worthwhile to study since the leptonic decays are
clean and rich from the theoretical point of view. They are clean
because of the fact that they are free from strong interactions.
On the other hand, they are rich since they exist at least in the
loop level. The lepton FV is tiny in the extended SM, with non
zero neutrino masses and the theoretical branching ratios (BRs) of
lepton flavor violating (LFV) decays are too small to reach the
experimental limits. Therefore, one needs to search more
fundamental models to pull the theoretical values to the current
experimental results. The radiative $l_i\rightarrow l_j\gamma$,
$i\neq j$ decays are among the LFV interactions and the current
experimental upper limits of the BRs of the decays $\mu\rightarrow
e\gamma$, $\tau\rightarrow e\gamma$ and $\tau\rightarrow
\mu\gamma$, read $1.2\times 10^{-11}$ \cite{Brooks}, $3.9\times
10^{-7}$ \cite{Hayasaka} and $1.1\times 10^{-6}\, (9.0\times
10^{-8}\,\mathbf{;}\,6.8\times 10^{-8}\mathbf{,}\,\, 90\% CL)$
\cite{Ahmed} (\cite{Roney}; \cite{Aubert}), respectively.
Furthermore, a new experiment at PSI has been described
\cite{Nicolo} in order to search the $\mu\rightarrow e \gamma$
decay and the aim is to reach to a sensitivity of BR$\sim
10^{-14}$. At present, this experiment (PSI-R-99-05 Experiment) is
still running in the MEG \cite{Yamada}.

The general two Higgs doublet model (2HDM) with the flavor
changing neutral currents (FCNCs) at tree level is the most
primitive model beyond the SM to enhance the BRs of the LFV decays
under consideration and the lepton FV is induced by the internal
new neutral Higgs bosons, $h^0$ and $A^0$. The Yukawa couplings
appear as free parameters and they can be determined by the
experimental data. These decays were examined in the framework of
the 2HDM in \cite{Iltan1,Iltan11, Diaz,
IltanExtrDim,IltanExtrDim1, IltanLFVSplit,IltanLFVFatSplit}.
Besides the theoretical calculations based on the 2HDM, they were
studied in the supersymmetric models
\cite{Barbieri1}-\cite{Barbieri7}, in a model independent way
\cite{Chang} and, recently, they were analyzed in the framework of
2HDM and the supersymmetric model \cite{Paradisi}.

The present work is devoted to the analysis of the radiative
$l_i\rightarrow l_j\gamma$, $i\neq j$ decays in the
2HDM\footnote{Here, we assume that the Cabibbo-Kobayashi-Maskawa
(CKM) type matrix in the leptonic sector does not exist, the
charged flavor changing (FC) interactions vanish and the lepton FV
comes from the internal new neutral Higgs bosons, $h^0$ and
$A^0$.}, including a single extra dimension, respecting the
Randall Sundrum scenario \cite{Rs1, Rs2}. The extra dimensions are
introduced to solve the hierarchy problem between weak and Planck
scales. The Randall Sundrum  model (the RS1 model) is based on the
non-factorizable geometry of  the extra dimension that the gravity
is localized on a 4D brane boundary, so called hidden (Planck)
brane, and the other fields, including the SM fields, live on
another 4D brane boundary, so called the visible (TeV) brane. The
difference of induced metrics on these boundaries are carried by a
warp factor that is an exponential function of the compactified
radius in the extra dimension and it connects two effective
scales, the Planck scale $M_{Pl}$ and the weak scale $m_W$. In the
case that some of the SM fields are accessible to the extra
dimension, the phenomenology becomes richer and there are various
work done respecting such scenarios in the literature
\cite{Goldberger}-\cite{Moreau2}. In \cite{Pamoral,Hewett} the
behavior of $U(1)$ gauge boson, living in the extra dimension of
the RS1 background, has been studied. In these works, it is
observed that the massless mode of the gauge field is not
localized in the extra dimension and its KK excitations have large
couplings to boundaries. Since it is necessary to push the visible
scale to energies greater than TeV for a perturbative theory, this
is not a phenomenologically favorable scenario. The brane
localized mass terms for scalar fields have been considered to get
small couplings of KK modes with the boundaries \cite{Pamoral2}.
In this case, these mass terms change the boundary conditions and
the zero mode localized solution is obtained. \cite{Batell} is
devoted to the localized $U(1)_Y$ gauge field with bulk and
boundary mass terms. In \cite{Huber4}, the KK excitations of W and
Z bosons at the LHC are studied. A further approach is to consider
that the fermions are also accessible to the extra dimension and
to explain the fermion mass hierarchy with the addition of Dirac
mass term to the Lagrangian. The bulk fermions are considered in
the RS1 background and the fermion mass hierarchy, coming from the
possible fermion field locations, is studied in \cite{Pamoral2,
Grossman, Huber, Huber2}. In \cite{Huber3} this hierarchy is
analyzed, by taking that the Higgs field has an exponential
profile around the TeV brane. \cite{Kogan} is devoted to an
extensive work on the bulk fields in various multi-brane models.
The quark and lepton FV, which is based on the different locations
of the fermion fields in the extra dimension, is extensively
studied in \cite{KAgashe, EBlechman} and in these works, it is
considered that the FV is carried by the Yukawa interactions,
coming from the SM Higgs-fermion-fermion vertices. In \cite{Pree},
the high precision measurements of top pair production at the ILC
is addressed  by considering that the fermions are localized in
the bulk of RS1 background. In recent works \cite{Moreau1,
Moreau2}, the various experimental FCNC constraints and the
electro weak precision tests for the location parameters of the
fermions in the extra dimension are discussed.

In this work, we study the LFV $l_i\rightarrow l_j\gamma$ decays
in the RS1 background with the assumption that the leptons and
gauge fields are accessible to the extra dimension. Furthermore,
we consider that the lepton fields are localized in the extra
dimension and their mass hierarchy is based on the lepton field
location in the extra dimension.

The paper is organized as follows: In Section 2, we present the
BRs of LFV interactions $l_i\rightarrow l_j\gamma$ in the 2HDM, by
considering that the lepton fields are localized in the extra
dimension of RS1 background. Section 3 is devoted to discussion
and our conclusions.
\section{LFV $l_i\rightarrow l_j\gamma$ decays in the Randall
Sundrum background with localized lepton fields}
The radiative LFV decays exist in the loop level and, therefore,
the theoretical expressions of their physical quantities contain
number of free parameters of the model used. The theoretical work
done on these decays ensures considerable information on the
parameters of the model, with the help of the experimental
measurements. In the framework of the SM, the numerical values of
the BRs of these decays are far from experimental limits. This
forces one to consider new models beyond the SM and the 2HDM, with
the FCNCs at tree level, is one of the candidate. The new neutral
Higgs bosons in the 2HDM switch on the LFV interactions and the
new Yukawa couplings, arising from lepton-lepton-scalar
interactions, play the main role in the calculation of the
physical quantities related to these decays.

Addition of the contributions coming from the extra dimension(s)
further enhances the BRs of the decays studied  and, in the
present work, we study these effects in the RS1 background, by
assuming that the charged leptons and the gauge bosons are
accessible to the extra dimension. In the RS1 model, the extra
dimension is curved and the metric reads
\begin{eqnarray}
ds^2=e^{-2\,\sigma}\,\eta_{\mu\nu}\,dx^\mu\,dx^\nu-dy^2\, ,
\label{metric1}
\end{eqnarray}
where $\sigma=k\,|y|$ and $k$ is the bulk curvature constant. Here
the exponential $e^{-k\,|y|}$, with $y=R\,|\theta|$,  is the warp
factor\footnote{The warp factor causes that all mass terms are
rescaled on the visible brane for $\theta=\pi$.} that is
responsible for the hierarchy and $R$ is the compactification
radius in the extra dimension that is compactified onto $S^1/Z_2$
orbifold, having two boundaries, the hidden (Planck) brane and the
visible (TeV) brane. In the RS1 background, the gravity is
considered to be localized on the hidden brane and to be extended
into the bulk with varying strength, on the other hand, the SM
fields live in the visible brane. By choosing opposite and equal
tensions on the hidden and visible branes, the low energy
effective theory has flat 4D spacetime, even if the 5D
cosmological constant is non vanishing. In the present work, we
assume that the gauge fields and the fermions are also accessible
to the extra dimension. The addition of Dirac mass term to the
lagrangian of bulk fermions results in the fermion localization in
the extra dimension \cite{Hisano, Hewett, Pamoral2, Huber4,
Grossman, Huber2,Huber3}. Since the fermions have two possible
transformation properties under the orbifold $Z_2$ symmetry,
$Z_2\psi=\pm\gamma_5 \psi$, the combination $\bar{\psi}\psi$ is
odd and, in order to construct the $Z_2$ invariant mass term, one
needs  $Z_2$ odd scalar field to be coupled. Therefore, the mass
term reads
\begin{eqnarray}
{\cal{S}}_m=-\int d^4x \int dy \,\sqrt{g}\,m(y)\,\bar{\psi}\psi
\label{massterm} \, ,
\end{eqnarray}
where $m(y)=m\frac{\sigma'(y)}{k}$ with
$\sigma'(y)=\frac{d\sigma}{dy}$ and
$g=Det[g_{MN}]=e^{-8\,\sigma}$, $M,N=0,1,... ,4$. By expanding the
bulk fermion as
\begin{eqnarray}
\psi(x^\mu,y)=\frac{1}{\sqrt{2\,\pi\,R}}\,\sum_{n=0}^\infty\,
\psi^{(n)}(x^\mu)\, e^{2\,\sigma}\, \chi_n(y) \label{psiKK} \, ,
\end{eqnarray}
and using the normalization
\begin{eqnarray}
\frac{1}{2\,\pi\,R}\,\int_{-\pi\,R}^{\pi\,R}\,dy\,e^\sigma \,
\chi_n(y)\,\chi_m(y)=\delta_{nm} \label{norm} \, ,
\end{eqnarray}
the Dirac equation is solved for the zero mode fermion  and it is
obtained as follows:
\begin{eqnarray}
\chi_0(y)=N_0\, e^{-r\,\sigma}\label{0mode} \, ,
\end{eqnarray}
where $r=m/k$ and the normalization constant $N_0$ reads
\begin{eqnarray}
N_0=\sqrt{\frac{k\,\pi \,R\,(1-2\,r)}{e^{k\,\pi
\,R\,(1-2\,r)}-1}}\label{norm0mode} \, .
\end{eqnarray}
The appropriately normalized solution
\begin{eqnarray}
\chi'_0(y)=e^{-\frac{\sigma}{2}}\,\chi_0(y) \label{0modep}
\end{eqnarray}
is localized in the extra dimension and the localization is
regulated by the parameter $r$. This solution is localized  near
the hidden (visible) brane if $r>\frac{1}{2}$ ($r<\frac{1}{2}$)
and it has a constant profile if $r=\frac{1}{2}$.

To construct the SM fermions, one needs to consider $SU(2)_L$
doublet $\psi_L$ and singlet $\psi_R$ with separate $Z_2$
projection conditions: $Z_2\psi_R=\gamma_5 \psi_R$ and
$Z_2\psi_L=-\gamma_5 \psi_L$ (see for example \cite{Hisano}). This
is the case that the zero mode fermions can get mass through the
$Z_2$ invariant left handed fermion-right handed fermion-Higgs
interaction, $\bar{\psi}_R\,\psi_L\, H $\footnote{Here, we
consider different location parameters $r$ for each left handed
and right handed part of different flavors.} and the location
parameters for fermion fields, different flavors and their left
and right handed parts, can be chosen so that this interaction
creates the current masses of fermions. With the assumption that
the Higgs field lives on the visible brane, the masses of fermions
are calculated by using the integral
\begin{eqnarray}
m_i=\frac{1}{2\,\pi\,R}\,\int_{-\pi\,R}^{\pi\,R}\,dy\,\lambda_5\,
\chi_{iL0}(y)\,\chi_{iR0}(y)\,<H>\,\delta(y-\pi\, R) \label{mi} \,
,
\end{eqnarray}
where $\lambda_5$ is the coupling in five dimensions and it can be
parametrized  in terms of the one in four dimensions, the
dimensionless coupling $\lambda$, $\lambda_5=\lambda/\sqrt{k}$.
Here the expectation value of the Higgs field $<H>$ reads
$<H>=v/\sqrt{k}$ where $v$ is the vacuum expectation
value\footnote{We take $v=0.043\,M_{Pl}$ to provide the measured
gauge boson masses \cite{Huber2} and choose $k\,R=10.83$ in order
to get the correct effective scale on the visible brane, i.e.,
$M_W=e^{-\pi\,k\,R}\, M_{pl}$ is of the order of TeV.}. Now, we
choose three different sets of location of charged lepton fields
in order to obtain the masses of different flavors.
\newpage
\begin{table}[h]
        \begin{center}
\begin{tabular}{|c|c|c|c|c|c|c|}
  \hline
    & SET I  & SET  II &  SET III \\
  \hline \hline
& $r_L$ \,\,\,\,\, $r_R$  & \,$r_L$ \,\,\,\,\,  $r_R$  & \,$r_L$  \,\,\,\,\,$r_R$  \\
\hline\hline
  e & 0.6710  \,\, 0.6710  & -0.4900 \,\, 0.8800 & -1.0000 \,\, 0.8860 \\ \hline
  $\mu$  & 0.5826 \,\,  0.5826  & -0.4900 \,\, 0.7160 & -1.0000 \,\, 0.7230 \\ \hline
  $\tau$ & 0.5273 \,\, 0.5273 & -0.4900 \,\,0.6249 & -1.0000 \,\,  0.6316 \\ \hline
  \hline
\end{tabular}
\end{center}
\caption{Three possible locations of charged lepton fields. Here
$r_L$ and $r_R$ are left handed and right handed lepton field
location parameters, respectively.} \label{set}
\end{table}
In Set I, the left and right handed fields of the same flavor have
the same location. In set II and III, we choose the left handed
charged lepton locations the same for each flavor and we estimate
the left handed charged lepton locations by respecting their
masses. Notice that, in the third set, we take the left handed
charged lepton locations nearer to the visible brane.

Now we would like to present the part of the action responsible
for the LFV interactions in 5 dimension:
\begin{eqnarray}
{\cal{S}}_{Y}= \int d^5x \sqrt{-g} \,\Bigg( \xi^{E}_{5\,
ij}\,\bar{l}_{i L} \phi_{2} E_{j R} + h.c. \Bigg)\, \delta(y-\pi
R) \,\,\, , \label{yukawalagrangian}
\end{eqnarray}
where $L$ and $R$ denote chiral projections $L(R)=1/2(1\mp
\gamma_5)$, $\phi_{2}$ is the new scalar doublet, $l_{i L}$ ($E_{j
R}$) are lepton doublets (singlets), $\xi^{E}_{5\,ij}$, with
family indices $i,j$ , are the Yukawa couplings in five dimensions
,which induce the FV interactions in the leptonic sector. We
assume that the Higgs doublet $\phi_1$, that lives on the visible
brane, has non-zero vacuum expectation value to ensure the
ordinary masses of the gauge fields and the fermions, however, the
second doublet, that lies also on the visible brane, has no vacuum
expectation value:
\begin{eqnarray}
\phi_{1}=\frac{1}{\sqrt{2}}\left[\left(\begin{array}{c c}
0\\v+H^{0}\end{array}\right)\; + \left(\begin{array}{c c} \sqrt{2}
\chi^{+}\\ i \chi^{0}\end{array}\right) \right]\, ;
\phi_{2}=\frac{1}{\sqrt{2}}\left(\begin{array}{c c} \sqrt{2}
H^{+}\\ H_1+i H_2 \end{array}\right) \,\, , \label{choice}
\end{eqnarray}
 and the vacuum expectation values are
\begin{eqnarray}
<\phi_{1}>=\frac{1}{\sqrt{2}}\left(\begin{array}{c c}
0\\v\end{array}\right) \,  \, ; <\phi_{2}>=0 \,\, .
\label{choice2}
\end{eqnarray}
By considering the gauge and $CP$ invariant Higgs potential which
spontaneously breaks  $SU(2)\times U(1)$ down to $U(1)$  as:
\begin{eqnarray}
V(\phi_1, \phi_2 )&=&c_1 (\phi_1^+ \phi_1-v^2/2)^2+ c_2 (\phi_2^+
\phi_2)^2 \nonumber \\ &+&  c_3 [(\phi_1^+ \phi_1-v^2/2)+ \phi_2^+
\phi_2]^2 + c_4 [(\phi_1^+ \phi_1)
(\phi_2^+ \phi_2)-(\phi_1^+ \phi_2)(\phi_2^+ \phi_1)] \nonumber \\
&+& c_5 [Re(\phi_1^+ \phi_2)]^2 + c_{6} [Im(\phi_1^+ \phi_2)]^2
+c_{7} \, , \label{potential}
\end{eqnarray}
with constants $c_i, \, i=1,...,7$, $H_1$ and $H_2$ are obtained
as the mass eigenstates $h^0$ and $A^0$ respectively, since no
mixing occurs between two CP-even neutral bosons $H^0$ and $h^0$
in the tree level. With the choice eq.(\ref{choice2}) and the
potential eq.(\ref{potential}), the SM particles can be collected
in the first doublet and the new particles in the second
one\footnote{In the case that the second Higgs doublet has also a
vacuum expectation value, there would be a mixing between the CP
even neutral Higgs bosons and there appears an additional free
parameter, the mixing angle, which should be restricted. In
general, this mixing is considered weak and, our choice of Higgs
doublets results in vanishing mixing between CP even neutral Higgs
bosons. This is the case that one free parameter is dropped
without loosing the crucial property of the model, namely, the FV
at tree level}. Since the lepton fields are accesible to extra
dimension, the lepton doublets $l_{i L}$ and singlets $E_{j R}$
are expanded as
%
%
\begin{eqnarray}
l_{i L}(x^\mu,y)&=&\frac{1}{\sqrt{2\,\pi\,R}}\, e^{2\,\sigma}\,
l_{i L}^{(0)}(x^\mu)\, \chi_{i\,L 0}(y)\nonumber
\\ &+& \frac{1}{\sqrt{2\,\pi\,R}}\,\sum_{n=1}^\infty\, e^{2\,\sigma}\,
\Bigg( l_{i L}^{(n)}(x^\mu)\, \chi^{l}_{i\,L n}(y)+l_{i
R}^{(n)}(x^\mu)\, \chi^{l}_{i\,R n}(y)\Bigg)\, , \nonumber
\\
E_{j R}(x^\mu,y)&=& \frac{1}{\sqrt{2\,\pi\,R}}\,
e^{2\,\sigma}\,E_{j R}^{(0)}(x^\mu)\, \chi_{j\,R 0}(y)\nonumber
\\ &+&\frac{1}{\sqrt{2\,\pi\,R}}\,\sum_{n=0}^\infty\,
e^{2\,\sigma}\,\Bigg( E_{j R}^{(n)}(x^\mu)\, \chi^E_{j\,R n}(y)+
E_{j L}^{(n)}(x^\mu)\, \chi^E_{j\,L n}(y)\Bigg) \label{leptonKK}
\, .
\end{eqnarray}
The zero mode leptons $\chi_{i\,L 0}(y)$ and  $\chi_{j\,R 0}(y)$
are given in eq. (\ref{0mode}) with the replacements $r\rightarrow
r_{iL}$ and $r\rightarrow r_{iR}$, respectively. For the effective
Yukawa coupling $\xi^{E}_{ij}$, we need to integrate out the
Yukawa interaction eq.(\ref{yukawalagrangian}) over the fifth
dimension by taking the zero mode lepton doublets, singlets, and
neutral Higgs fields $S=h^0,A^0$:
\begin{eqnarray}
V^0_{RL\,ij}&=&\frac{\xi^{E}_{5\,
ij}}{2\,\pi\,R}\,\int_{-\pi\,R}^{\pi\,R}\,dy\,
\chi_{iR0}(y)\,\chi_{jL0}(y)\,\delta(y-\pi R) \nonumber \\ &=&
\frac{e^{-k\,\pi\,R\,(r_{iR}+r_{jL})}\,k\,
\sqrt{(1-2\,r_{iR})\,(1-2\,r_{jL})}}
{\sqrt{(e^{k\,\pi\,R\,(1-2\,r_{iR})}-1)\,
(e^{k\,\pi\,R\,(1-2\,r_{jL})}-1)}}\, \xi^{E}_{5\,
ij}\label{yukawa0LR} \, .
\end{eqnarray}
At this stage, we consider two different scenarios. At first, we
assume that the hierarchy of new Yukawa couplings comes from the
lepton field localization and, by considering that the left and
right handed fields of the same flavor have the same location (Set
I), we construct the Yukawa coupling $\xi^{E}_{ij}$ and its
hermitian conjugate as follows:
\begin{eqnarray}
\xi^{E}_{ij}\,\Big((\xi^{E}_{ij})^\dagger\Big)=
\frac{V^0_{RL\,(LR)\,ij}}{V^0_{RL\,(LR)\,\mu\tau}} \, \alpha\,\, .
\label{set1Yukawa}
\end{eqnarray}
Here, we fix the new neutral Higgs $S-\tau-\mu$ coupling (renamed
as $\alpha$) in four dimensions, and we estimate the other
couplings, respecting the possible locations of the charged
leptons which are calculated by considering their mass
hierarchy\footnote{This case is similar to Sher anzats \cite{Sher}
that the couplings are defined as $\xi^{E}_{ij}\sim
\sqrt{m_i\,m_j}$ }. The idea behind is to consider the $\tau-\mu$
coupling in four dimensions as non-zero in the case of $S$-charged
lepton-charged lepton interaction (zero for the case of SM
Higgs-charged lepton-charged lepton interaction since the off
diagonal entries of the SM Higgs-charged lepton-charged lepton
couplings in five dimensions vanish, in our choice) and it is
fixed by embedding the contribution coming from the mixing due to
the different locations of L(R) handed muon-R(L) handed
tau\footnote{Notice that we choose $c_L=c_R$ for this case.} into
the number $\alpha$, with the choice of flavor dependent coupling
in five dimensions. In other words, the flavor dependence due to
the different locations of L(R) handed muon, R(L) handed tau is
compensated by the flavor dependence of the coupling in five
dimensions and, at the end, we get a fixed number $\alpha$. As a
second scenario, we embed the vertex factor $V^0_{RL\,(LR)\,ij}$
into the coupling
$\xi^{E}_{ij}\,\Big((\xi^{E}_{ij})^\dagger\Big)$. In this case,
similar to the previous one, the coupling $\xi^{E}_{5\, ij}$ in
five dimension is flavor dependent and it is regulated in such a
way that the overall quantity $V^0_{RL\,(LR)\,ij}$ is pointed to
the chosen numerical value of
$\xi^{E}_{ij}\,\Big((\xi^{E}_{ij})^\dagger\Big)$. In this
scenario, the hierarchy of new Yukawa couplings is not related to
the lepton field locations.
%
%
Furthermore, the strengths of $S$-KK mode charged lepton-charged
lepton couplings are regulated by the locations of the lepton
fields. In our second scenario, we use two different sets, Set II
and Set III, for the locations of the lepton fields. Notice that
in both scenarios, the FV is carried by the new Yukawa couplings
in four dimensions and the effect of extra dimension is the
enhancement in the physical quantities of the processes
studied.\footnote{These scenarios are different than the ones
presented in the works \cite{KAgashe, EBlechman} where the sources
of the FV are explicitly the parameters related to the different
locations of the fermion fields in the extra dimension. In these
works, it is considered that the FV is carried by the Yukawa
interactions, coming from the SM Higgs-fermion-fermion vertices
with the help of the coupling in five dimensions, including
non-zero off diagonal entries.}

\begin{figure}[htb]
\vskip -6.0truein \centering
hspace[6cm]hspace[6cm]hspace[1cm]\epsfxsize=8.8in
\leavevmode\epsffile{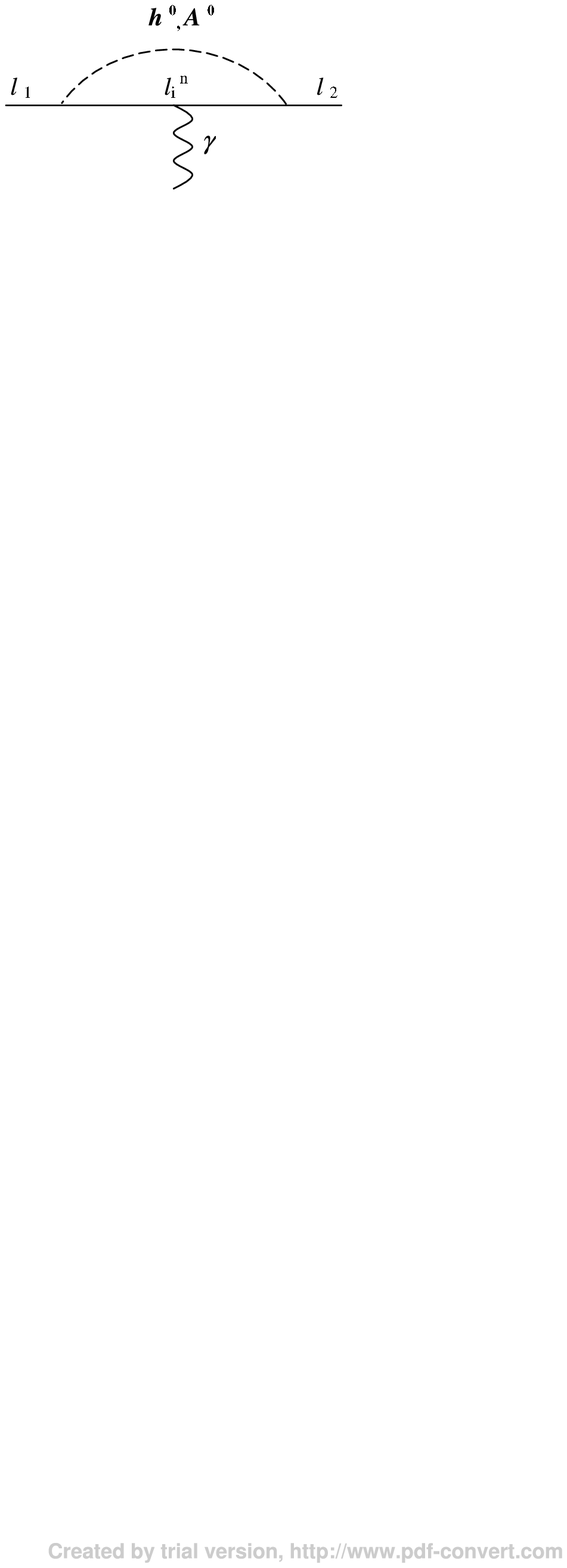} \vskip -11.0truein \caption[]{One
loop diagrams contribute to $l_1\rightarrow l_2 \gamma$ decay  due
to the zero mode (KK mode) leptons in the 2HDM.} \label{fig1}
\end{figure}

Since the radiative LFV decays under consideration exist at least
in the one loop level, the $S$-charged lepton-KK charged lepton
vertices arise and these vertex factors should be taken into
account. The $Z_2$ the projection condition $Z_2\psi=-\gamma_5
\psi$, used in constructing the left handed fields on the branes,
results in that the left handed zero mode appears, the left
(right) handed KK modes appear (disappear) on the branes, with the
boundary conditions due to the Dirac mass term in the action
eq.(\ref{massterm}):
\begin{eqnarray}
\Big(\frac{d}{dy}-m \Big)\,\chi^l_{iLn}(y_0)=0 \nonumber \\
\chi^l_{iRn}(y_0)=0 \, , \label{nLbound}
\end{eqnarray}
where $y_0=0$ or $\pi\,R$. Using the Dirac equation for KK mode
leptons one gets the left handed lepton $\chi^l_{i\,L n}(y)$ that
lives on the visible brane as
\begin{eqnarray}
\chi^l_{iLn}(y)=N_{Ln}\, e^{\sigma/2} \Bigg( J_{\frac{1}{2}-r}
(e^{\sigma}\,x_{nL})+c_L\, Y_{\frac{1}{2}-r}
(e^{\sigma}\,x_{nL})\Bigg)\label{nLmode} \, ,
\end{eqnarray}
with the constant
\begin{eqnarray}
c_L=-\frac{J_{-r-\frac{1}{2}} (x_{nL})}{Y_{-r-\frac{1}{2}}
(x_{nL})} \, .  \label{cL}
\end{eqnarray}
Here, $N_{Ln}$ is the normalization constant and
$x_{nL}=\frac{m_{Ln}}{k}$. The functions $J_\beta(w)$ and
$Y_\beta(w)$ appearing in eq.(\ref{nLmode}) are the Bessel
function of the first kind and of the second kind, respectively.
On the other hand, the $Z_2$ projection condition
$Z_2\psi=\gamma_5 \psi$ is used in order to construct the right
handed fields on the branes and this ensures that the right handed
zero mode appears, the right (left) handed KK modes appear
(disappear) on the branes with the boundary conditions:
\begin{eqnarray}
\Big(\frac{d}{dy}+m \Big)\,\chi^E_{iRn}(y_0)=0 \nonumber \\
\chi^E_{iLn}(y_0)=0 \, . \label{nRbound}
\end{eqnarray}
Again, using the Dirac equation for KK mode leptons, one gets the
right handed lepton $\chi^E_{i\,R n}(y)$ that lives on the visible
brane as
\begin{eqnarray}
\chi^E_{iRn}(y)=N_{Rn}\, e^{\sigma/2} \Bigg( J_{\frac{1}{2}+r}
(e^{\sigma}\,x_{nR})+c_R\, Y_{\frac{1}{2}+r}
(e^{\sigma}\,x_{nR})\Bigg)\label{nRmode} \, ,
\end{eqnarray}
with
\begin{eqnarray}
c_R=-\frac{J_{r-\frac{1}{2}} (x_{nR})}{Y_{r-\frac{1}{2}} (x_{nR})}
\, , \label{cR}
\end{eqnarray}
where $N_{Rn}$ is the normalization constant and
$x_{nR}=\frac{m_{Rn}}{k}$. Notice that the constant $c_L$, the
$n^{th}$ KK mode mass $m_{Ln}$ in eq.(\ref{nLmode}) and the
constant $c_R$, the $n^{th}$ KK mode mass $m_{Rn}$ in
eq.(\ref{nRmode}) are obtained by using the boundary conditions
eq.(\ref{nLbound}) and eq.(\ref{nRbound}), respectively. For
$m_{L(R)n}\ll k$ and $kR\gg 1$ they are approximated as:
\begin{eqnarray}
m_{Ln}&\simeq&
k\,\pi\,\Big(n-\frac{\frac{1}{2}-r}{2}+\frac{1}{4}\Big)\,
e^{-\pi\,k\,R} \nonumber \, ,\\
m_{Rn} &\simeq&
k\,\pi\,\Big(n-\frac{\frac{1}{2}+r}{2}+\frac{1}{4}\Big)\,e^{-\pi\,k\,R}
\,\,\,\,\,\,\,\, \mbox{for $r<0.5$} \nonumber \, ,\\
m_{Rn}&\simeq&
k\,\pi\,\Big(n+\frac{\frac{1}{2}+r}{2}-\frac{3}{4}\Big)\,e^{-\pi\,k\,R}
\,\,\,\,\,\,\,\, \mbox{for $r>0.5$}\label{mnLR} \, .
\end{eqnarray}
Now, we are ready to calculate the $S$-charged lepton-KK charged
lepton vertex factors,  $V^n_{LR\,(RL)\,ij}$, that appear in the
loop diagram (see Fig.\ref{fig1}). The procedure is the
integration of the Yukawa interaction eq.(\ref{yukawalagrangian})
over the fifth dimension by taking the $S$-zero mode lepton
singlet (doublet)-KK mode lepton doublet (singlet)
\begin{eqnarray}
V^n_{RL\,(LR)\,ij}=\frac{\xi^{E}_{5\, ij}\,\Big((\xi^{E}_{5\,
ij})^{\dagger}\Big) }{2\,\pi\,R}\,\int_{-\pi\,R}^{\pi\,R}\,dy\,\,
\chi_{iR0\,(iL0)}(y)\,\, \chi_{jLn\,(jRn)}(y)\,\,\delta(y-\pi\,
R)\label{intyukawa0KK} \, ,
\end{eqnarray}
and we get
\begin{eqnarray}
V^n_{RL\,\,ij}&=&\frac{ N_{Ln}\, e^{k \pi R\,(1/2-r_{iR})}\,
\Bigg( J_{\frac{1}{2}-r_{jL}} (e^{k \pi R}\,x_{nL})+c_L\,
Y_{\frac{1}{2}-r_{jL}}(e^{k \pi R}\,x_{nL})\Bigg) } {\pi R \sqrt{
\frac{e^{k\,\pi \,R\,(1-2\,r_{iR})}-1}{k\,\pi
\,R\,(1-2\,r_{iR})}}}\,\xi^{E}_{5\, ij}
\nonumber \, , \\
V^n_{LR\,\,ij}&=& \frac{N_{Rn}\, e^{k \pi R\,(1/2-r_{iL})}\,
\Bigg( J_{\frac{1}{2}+r_{jR}} (e^{k \pi R}\,x_{nR})+c_R\,
Y_{\frac{1}{2}+r_{jR}} (e^{k \pi R}\,x_{nR})\Bigg)}{\pi R
\sqrt{\frac{e^{k\,\pi \,R\,(1-2\,r_{iL})}-1}{k\,\pi
\,R\,(1-2\,r_{iL})}}}\, (\xi^{E}_{5\,
ij})^{\dagger}\label{yukawa0LKKR} \, ,
\end{eqnarray}
where $c_L$ ($c_R$) is given in eq.(\ref{cL}) (eq.(\ref{cR})) with
the replacements $r\rightarrow r_{jL}$ ($r\rightarrow r_{jR}$). In
the first scenario that the hierarchy of new Yukawa couplings
comes from the lepton field location, the effective $S$-zero mode
lepton singlet (doublet)-KK mode lepton doublet (singlet) coupling
$\xi^{E\, n}_{ij}\,(\xi^{E\, n}_{ij})^\dagger$ reads
\begin{eqnarray}
\xi^{E\,n}_{ij}\,\Big((\xi^{E\,n}_{ij})^\dagger\Big)=
\frac{V^n_{RL\,(LR)\,ij}}{V^0_{RL\,(LR)\,\mu\tau}} \,\alpha \,\, ,
\label{set1Yukawan}
\end{eqnarray}
and, in the second scenario, we get
\begin{eqnarray}
\xi^{E\, n}_{ij}\,\Big((\xi^{E\,n}_{ij})^\dagger\Big)=
\frac{V^n_{RL\,(LR)\,ij}}{V^0_{RL\,(LR)\,ij}}\, \xi^{E}_{ij} \,\,
. \label{set2Yukawan}
\end{eqnarray}
At this stage, we present the BRs of the LFV $\mu\rightarrow
e\gamma$, $\tau\rightarrow e\gamma$ and $\tau\rightarrow
\mu\gamma$ decays. The existence of decays at least in the one
loop level brings the logarithmic divergences in the calculations
and we eliminate them by using the on-shell renormalization
scheme. In this scheme, the self energy diagrams for on-shell
leptons vanish since they can be written as $
\sum(p)=(\hat{p}-m_{l_1})\bar{\sum}(p) (\hat{p}-m_{l_2})\, , $ and
the vertex diagram in Fig.\ref{fig1} gives non-zero
contribution.\footnote{This is the case that the divergences can
be eliminated by introducing a counter term $V^{C}_{\mu}$ with the
relation $V^{Ren}_{\mu}=V^{0}_{\mu}+V^{C}_{\mu} \, , $ where
$V^{Ren}_{\mu}$ ($V^{0}_{\mu}$) is the renormalized (bare) vertex
and by using the gauge invariance: $k^{\mu} V^{Ren}_{\mu}=0$.
Here, $k^\mu$ is the four momentum vector of the outgoing
photon.}. Taking only tau lepton for the internal line
\footnote{We take into account only the internal tau lepton
contribution since, in the second scenario, we respect the idea
that the couplings $\bar{\xi}^{E}_{N, ij}$ ($i,j=e,\mu$), are
small compared to $\bar{\xi}^{E}_{N,\tau\, i}$ $(i=e,\mu,\tau)$,
due to the possible proportionality of them to the masses of
leptons under consideration in the vertices. In the first
scenario, this idea is considered automatically.  Here, we use the
dimensionful coupling $\bar{\xi}^{E}_{N,ij}$ with the definition
$\xi^{E}_{N,ij}=\sqrt{\frac{4\, G_F}{\sqrt{2}}}\,
\bar{\xi}^{E}_{N,ij}$ where N denotes the word "neutral".}, the
decay width $\Gamma$ reads
\begin{eqnarray}
\Gamma (l_1\rightarrow l_2\gamma)=c_1(|A_1|^2+|A_2|^2)\,\, ,
\label{DWmuegam}
\end{eqnarray}
where
\begin{eqnarray}
A_1&=&Q_{\tau} \frac{1}{48\,m_{\tau}^2} \Bigg \{ 6\,m_\tau\,
\bar{\xi}^{E}_{N,\tau l_2}\, \bar{\xi}^{E}_{N,l_1\tau}\, \Big( F
(z_{h^0})-F (z_{A^0}) \Big)+ \sum_{n=1}^{\infty} m_{l_1}\,\Bigg(
\bar{\xi}^{E}_{N,l_2\tau}\, \bar{\xi}^{E}_{N,
l_1\tau}\, \Big(G (z_{h^0})+G(z_{A^0})\Big) \nonumber \\
&+& \sum_{n=1}^{\infty}\,\frac{m_{\tau}^2}{m^2_{nR}}
\,(\bar{\xi}^{E\,n}_{N, l_2\tau})^\dagger\,
(\bar{\xi}^{E\,n}_{N,l_1\tau})^\dagger\, \Big(G (z_{nR, h^0})+G
(z_{nR, A^0})\Big) \Bigg) \Bigg\}
\nonumber \,\, , \\
A_2&=&Q_{\tau} \frac{1}{48\,m_{\tau}^2} \Bigg \{6\,m_\tau\,
\bar{\xi}^{E}_{N, l_2 \tau}\, \bar{\xi}^{E}_{N,\tau l_1}\, \Big(F
(z_{h^0})-F(z_{A^0})\Big)+ m_{l_1}\,\Bigg(
\bar{\xi}^{E}_{N,l_2\tau}\, \bar{\xi}^{E}_{N,l_1
\tau}\, \Big( G (z_{h^0})+G (z_{A^0})\Big) \nonumber \\
&+& \sum_{n=1}^{\infty}\,\,\frac{m_{\tau}^2}{m^2_{nL}}
\bar{\xi}^{E\,n}_{N,l_2\tau}\, \bar{\xi}^{E\,n}_{N,l_1 \tau}\,
\Big(G (z_{nL, h^0})+ G (z_{nL, A^0}) \Big) \Bigg) \Bigg\}
 \,\, , \label{A1A2}
\end{eqnarray}
where $l_1\,(l_2)=\tau;\mu\,(\mu$ or $e; e)$ and the functions $F
(w)$, $G (w)$ are given by
\begin{eqnarray}
F (w)&=&\frac{w\,(3-4\,w+w^2+2\,ln\,w)}{(-1+w)^3} \, , \nonumber \\
G (w)&=&\frac{w\,(2+3\,w-6\,w^2+w^3+ 6\,w\,ln\,w)}{(-1+w)^4} \,\,
. \label{functions2}
\end{eqnarray}
Here $c_1=\frac{G_F^2 \alpha_{em} m^3_{l_1}}{32 \pi^4}$, $A_1$
($A_2$) is the left (right) chiral amplitude,
$z_{S}=\frac{m^2_{\tau}}{m^2_{S}}$, $z_{nL(nR),
S}=\frac{m^2_{nL\,(nR)}}{m^2_{S}}$ with left (right) handed
internal lepton KK mode mass $m_{nL\,(nR)}$ (eq.(\ref{mnLR})),
$Q_{\tau}$ is the charge of tau lepton. Notice that we take the
Yukawa couplings real and ignore the mass $m_e$ $(m_\mu)$ for
$\mu\rightarrow e \gamma$; $\tau\rightarrow e \gamma$
($\tau\rightarrow \mu \gamma$) decays.
%
\section{Discussion}
The radiative LFV $l_i\rightarrow l_j\gamma$ interactions exist at
least in the one loop level in the 2HDM, with the help of the
Yukawa interactions coming from  lepton-lepton-$S$ vertices. The
Yukawa couplings are free parameters which should be restricted by
using the current and forthcoming experiments. In the present
work, we study these decays in the RS1 background with the
assumption that the leptons are accessible to the extra dimension.
Furthermore, we choose the idea that the lepton fields are
localized in the extra dimension by considering a Dirac mass term
$m_l=r \sigma'$ with $\sigma=k\,|y|$ (eq.(\ref{massterm})). This
choice results in that the right and left handed lepton zero
modes, the SM leptons, have exponential profiles in the extra
dimension, given in eq.(\ref{0mode}), and their different
locations can explain different flavor mass
hierarchy.\footnote{There is another scenario, so called the split
fermion scenario, where the mass hierarchy comes from the
different locations of various flavors and their left and right
handed parts of lepton fields which have Gaussian profiles in the
flat extra dimensions. In this case, the localizations of fermions
can be obtained by coupling them to a scalar field having kink
profile, and the Gaussian profiles of leptons are obtained as an
approximation (see \cite{Hamed, Mirabelli} for details).} Here we
choose three different set of locations of charged leptons in
order to obtain the masses  of different flavors. Since we
consider that the leptons are accessible to the bulk, the gauge
sector of the model should live also in the extra dimension and
their KK modes appear after the compactification of the extra
dimension. The different fermion locations can induce additional
FCNC effects at tree level due to the couplings of neutral gauge
KK modes-leptons and they should be suppressed even for low KK
masses, by choosing the location parameters $c_L$ ($c_R$)
appropriately. In the set of location parameters we use (Table
\ref{set}), we verify the various experimental FCNC constraints
with KK neutral gauge boson masses as low as few TeVs (see the
similar the set of location parameters and the discussion given in
\cite{Moreau1, Moreau2}). On the other hand, in our scenario,
since the FV is carried by the new Yukawa couplings which are
fixed to an appropriate number, respecting the current
measurements, the location parameters of leptons are responsible
for the lepton mass hierarcies and the strengths of $S$-zero mode
charged lepton-KK mode charged leptonvertices. Therefore, they do
not play a crucial role in the existence of the FV in the tree
level and the constraints coming from various LFV processes become
more relaxed.

In the first set (Table \ref{set}), we consider the left and right
handed fields having the same location in the extra dimension.
This is the case that we construct the hierarchy of new Yukawa
couplings by fixing the $\tau-\mu$ coupling
$\bar{\xi}^{E}_{N,\tau\,\mu}$, which is renamed as $\alpha$ in the
text. In the second and third set, we choose the left handed
charged lepton locations as the same for each flavor and we
estimate the right handed ones by respecting the charged lepton
masses. In the third set, we take the left handed charged lepton
locations near to the visible brane and we observe that the KK
mode charged lepton contribution to the BRs of the decays studied
increases. The reason behind this enhancement is that the KK mode
couplings to the new Higgs scalars, living on the visible brane,
become stronger if the left handed lepton field is near to this
brane. For the second and third set of locations, we embed the
results of integrals of the $S$-zero mode charged lepton-zero mode
charged lepton couplings into the Yukawa couplings of
corresponding vertices. In this case, we choose the Yukawa
couplings such that $\bar{\xi}^{E}_{N,ij},\, i,j=e,\mu $ are
smaller compared to $\bar{\xi}^{E}_{N,\tau\, i}\, i=e,\mu,\tau$,
since latter ones contain heavy flavor and we assume that, in four
dimensions, the couplings $\bar{\xi}^{E}_{N,ij}$ is symmetric with
respect to the indices $i$ and $j$. Furthermore, the curvature
parameter $k$ and the compactification radius $R$ are the
additional free parameters of the theory. Here we take
$k\,R=10.83$ and consider in the region $10^{17}\,(GeV)\leq k \leq
10^{18} \,(GeV)$ (see the discussion in section 2 and
\cite{Huber2}). Throughout our calculations we use the input
values given in Table (\ref{input}).
\begin{table}[h]
        \begin{center}
        \begin{tabular}{|l|l|}
        \hline
        \multicolumn{1}{|c|}{Parameter} &
                \multicolumn{1}{|c|}{Value}     \\
        \hline \hline
        $m_{\mu}$                   & $0.106$ (GeV) \\
        $m_{\tau}$                  & $1.78$ (GeV) \\
        $m_{h^0}$           & $100$   (GeV)  \\
        $m_{A^0}$           & $200$   (GeV)  \\
        $G_F$             & $1.16637 10^{-5} (GeV^{-2})$  \\
        \hline
        \end{tabular}
        \end{center}
\caption{The values of the input parameters used in the numerical
          calculations.}
\label{input}
\end{table}

\begin{figure}[htb]
\vskip -3.0truein \centering \epsfxsize=6.8in
\leavevmode\epsffile{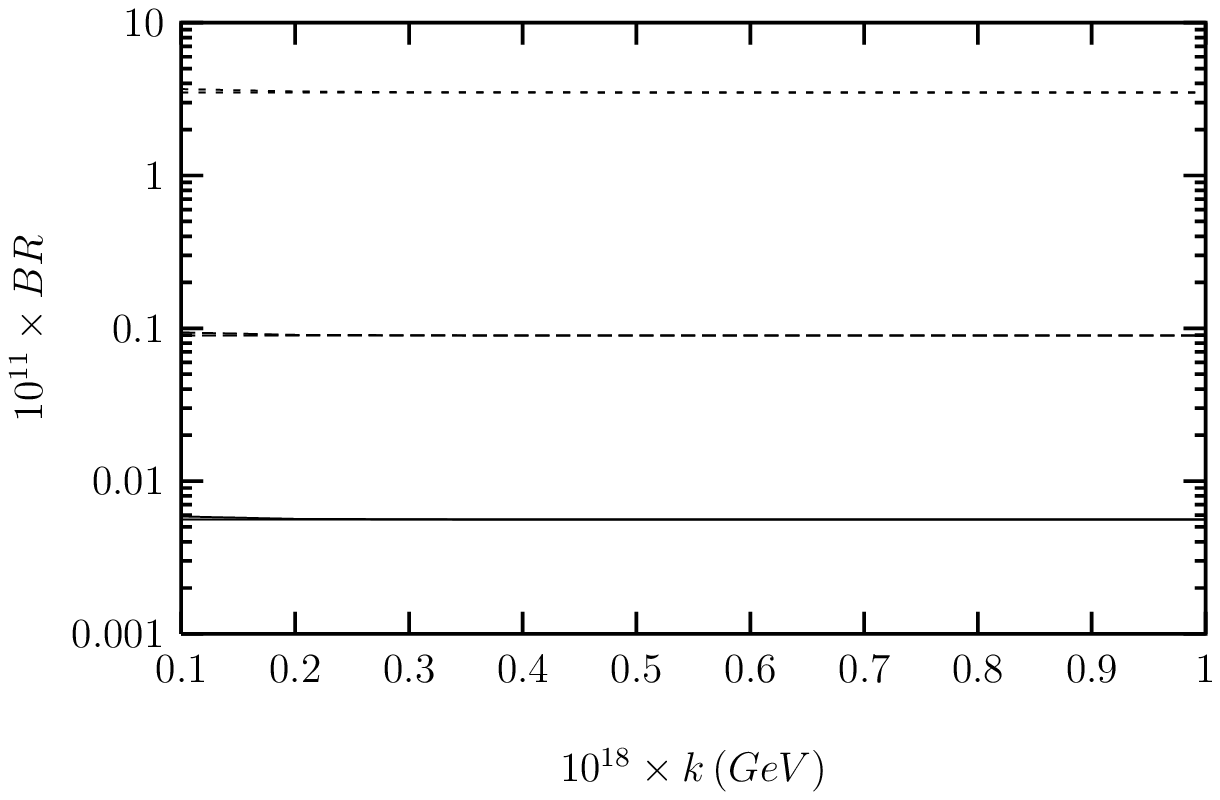} \vskip -3.0truein \caption[]{
The parameter $k$ dependence of the BR of the  LFV $\mu\rightarrow
e \gamma$ decay for the location Set I and different values of the
coupling $\alpha$. The solid (dashed, small dashed) line-curve
represents the BR for $\alpha=0.1\,(0.2, 0.5)\, GeV$ without-with
the lepton KK modes.} \label{BRmuegamkI}
\end{figure}

\begin{figure}[htb]
\vskip -3.0truein \centering \epsfxsize=6.8in
\leavevmode\epsffile{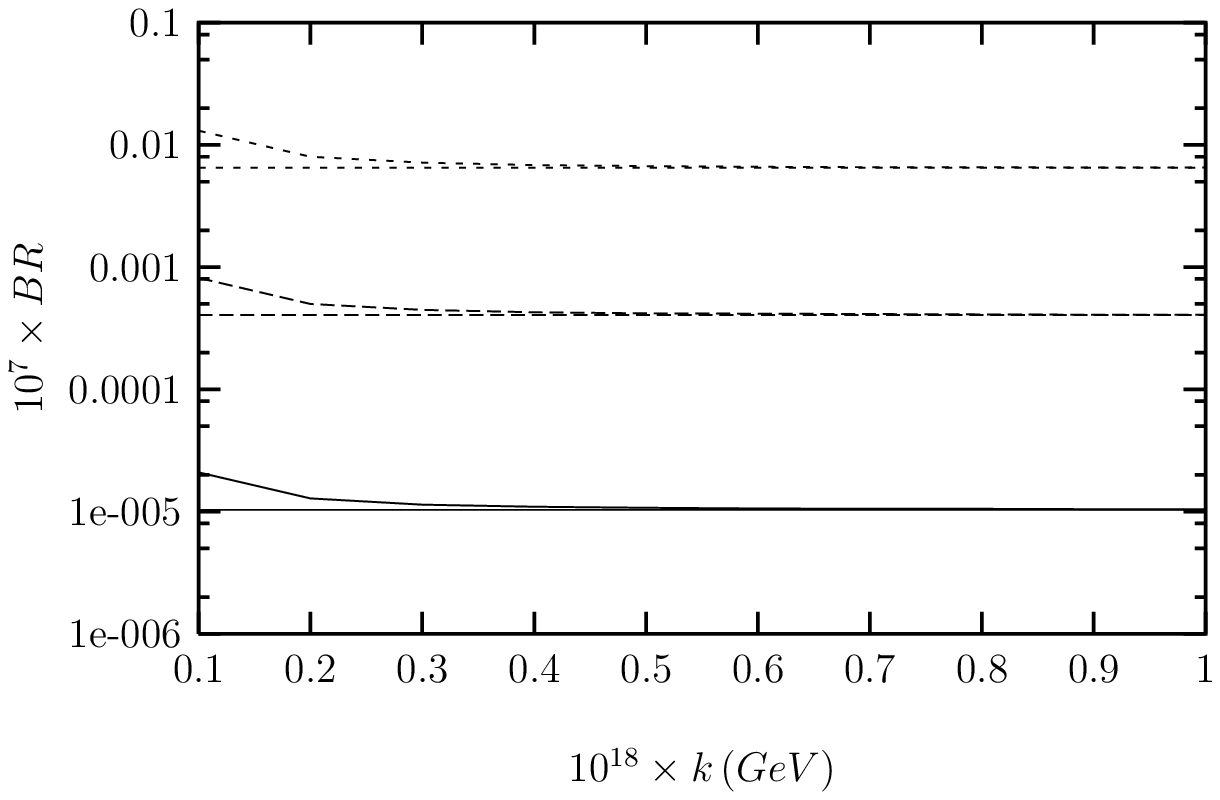} \vskip -3.0truein \caption[]{
The same as Fig.\ref{BRmuegamkI} but for  $\tau\rightarrow e
\gamma$ decay and  $\alpha=2\,(5, 10)\, GeV$.} \label{BRtauegamkI}
\end{figure}

\begin{figure}[htb]
\vskip -3.0truein \centering \epsfxsize=6.8in
\leavevmode\epsffile{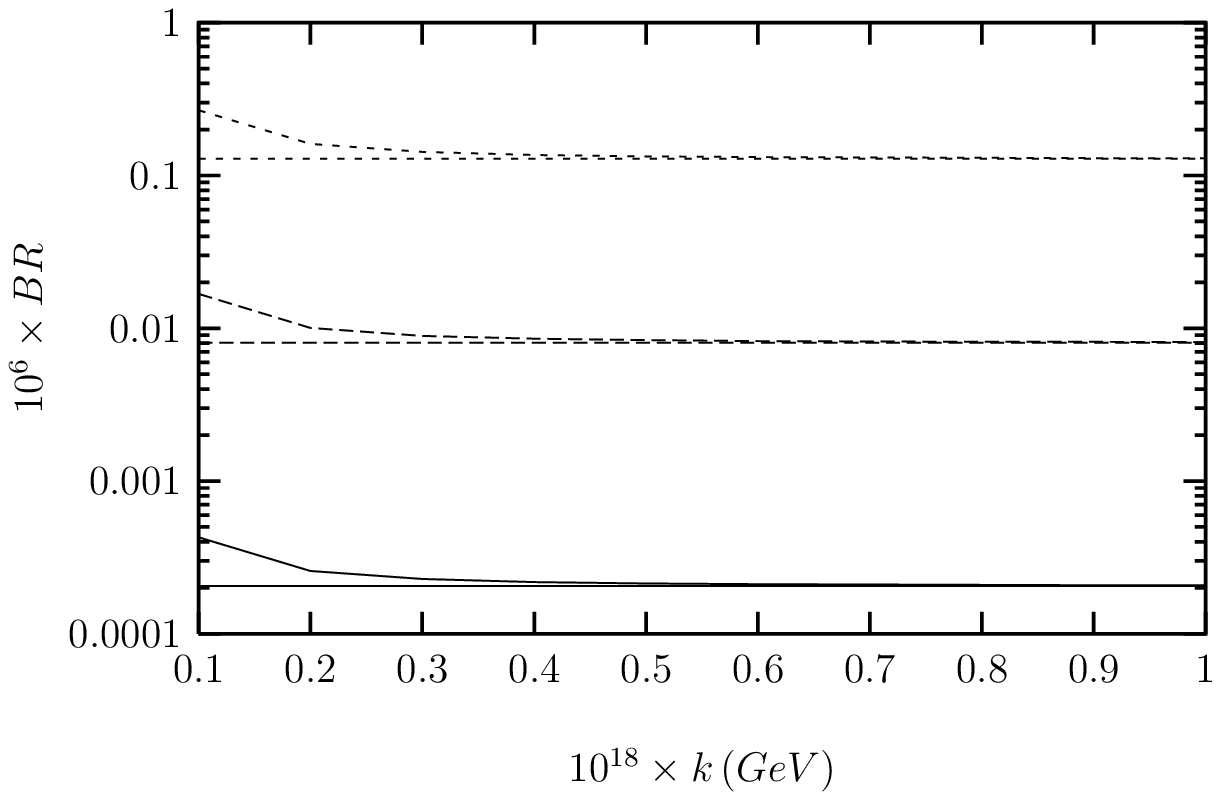} \vskip -3.0truein
\caption[]{ The same as Fig.\ref{BRtauegamkI} but for
$\tau\rightarrow \mu \gamma$ decay.} \label{BRtaumugamkI}
\end{figure}

In Figs.\ref{BRmuegamkI}-\ref{BRtauegamkI}-\ref{BRtaumugamkI}, we
present the parameter $k$ dependence of the BR of the  LFV decay
$\mu\rightarrow e \gamma$ - $\tau\rightarrow e \gamma$ -
$\tau\rightarrow \mu \gamma$ for location Set I, for different
values of the coupling $\alpha$. In Fig. \ref{BRmuegamkI}, the
solid (dashed, small dashed) line-curve represents the BR for
$\alpha=0.1\,(0.2, 0.5)\, GeV$ without -with the lepton KK modes.
In Figs.\ref{BRtauegamkI} and \ref{BRtaumugamkI}, the solid
(dashed, small dashed) line-curve represents the BR for
$\alpha=2\,(5, 10)\, GeV$ without -with the lepton KK modes,
similar to the previous figure. Fig.\ref{BRmuegamkI} shows that
the BR ($\mu\rightarrow e \gamma$) is at the order of the
magnitude of $10^{-13}\, (10^{-12}, 10^{-11})$ for $\alpha=0.1\,
(0.2, 0.5)\, GeV$ and it is not sensitive to the parameter $k$ in
the given range. Here, the enhancement due to the additional
contribution coming from internal lepton KK modes is negligible.
The BR ($\tau\rightarrow e \gamma$) - BR ($\tau\rightarrow
\mu\gamma$) is at the order of the magnitude of $10^{-12}\,
(10^{-10}, 10^{-9})$ - $10^{-10}\, (10^{-8}, 10^{-7})$ for
$\alpha=2\, (5, 10)\, GeV$.  For these decays the BR slightly
increases for the small values of $k$, $k \sim 10^{17}$ and the
enhancement in the BR due to the lepton KK mode contributions is
almost $100\%$ - more than $100\%$. To obtain the BR of the decay
$\mu\rightarrow e \gamma$ in the range $10^{-14}\leq BR\leq
10^{-11}$, the coupling $\alpha$ should lie in the region $0.1\leq
\alpha \leq 0.5$. On the other hand, this coupling  should be  in
the region $5\leq \alpha \leq 10$ for the case that the BR of the
decay $\tau\rightarrow e \gamma$ ($\tau\rightarrow \mu \gamma$)
with internal lepton KK modes has the numerical value lying in the
range $10^{-10}\leq BR\leq 10^{-9}$ ($10^{-8}\leq BR\leq
10^{-6}$). These numerical values show that the idea of hierarchy
of Yukawa couplings based on the locations of charged leptons in
the bulk of RS1 background can not explain the present
experimental results of these LFV decays and this stimulates to
perform more accurate measurements.

\begin{figure}[htb]
\vskip -3.0truein \centering \epsfxsize=6.8in
\leavevmode\epsffile{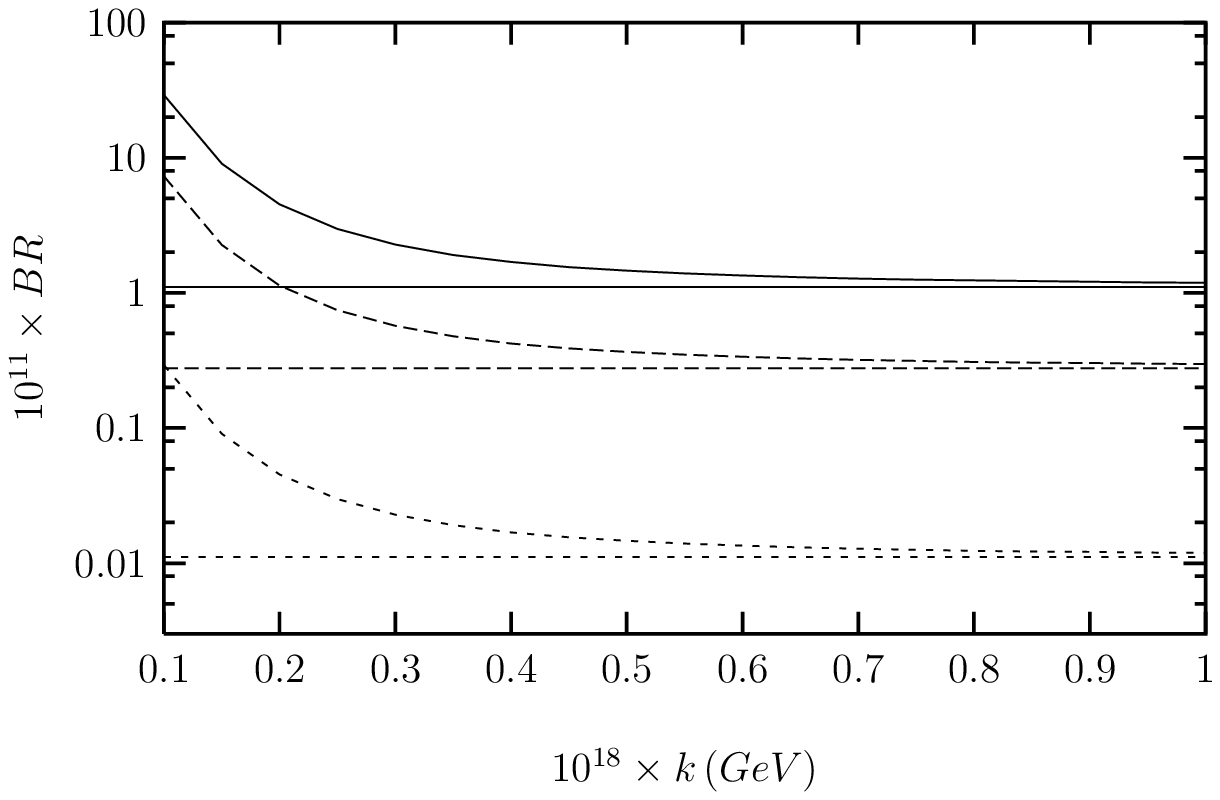} \vskip -3.0truein \caption[]{
The parameter $k$ dependence of the BR of the  LFV $\mu\rightarrow
e \gamma$ decay for the location Set II, for
$\bar{\xi}^{E}_{N,\tau\mu}=1\,GeV$ and for different values of the
coupling $\bar{\xi}^{E}_{N,\tau e}$. Here, the solid (dashed,
small dashed) line-curve represents the BR for and
$\bar{\xi}^{E}_{N,\tau e}=0.01\,(0.005, 0.001)\, GeV$ without
-with the lepton KK modes.} \label{BRmuegamkII}
\end{figure}

\begin{figure}[htb]
\vskip -3.0truein \centering \epsfxsize=6.8in
\leavevmode\epsffile{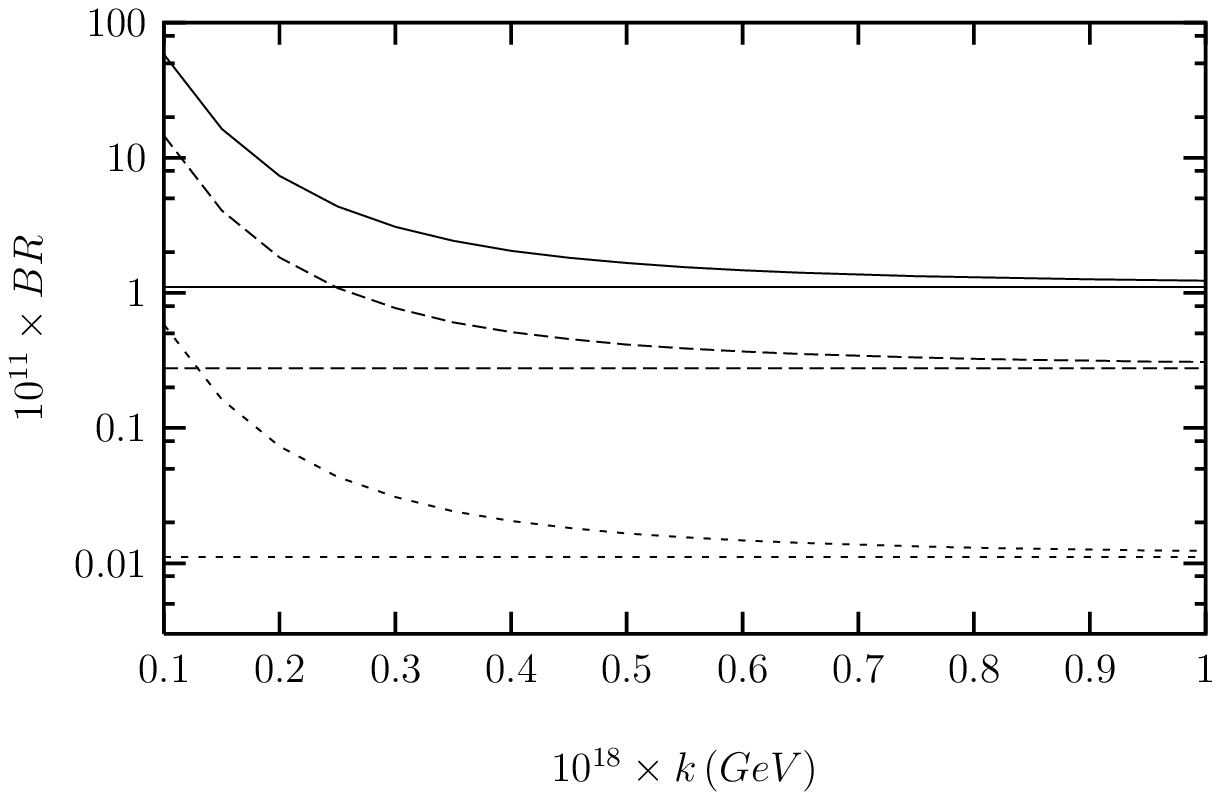} \vskip -3.0truein
\caption[]{ The same as Fig. \ref{BRmuegamkII} but for the
location Set III.} \label{BRmuegamkIId1}
\end{figure}

Figs. \ref{BRmuegamkII}-\ref{BRmuegamkIId1} are devoted to the
parameter $k$ dependence of the BR of the LFV $\mu\rightarrow e
\gamma$ decay for lepton location set II-III, for
$\bar{\xi}^{E}_{N,\tau\mu}=1GeV$ and for different values of the
coupling $\bar{\xi}^{E}_{N,\tau e}$. Here, the solid (dashed,
small dashed) line-curve represents the BR for
$\bar{\xi}^{E}_{N,\tau e}=0.01\,(0.005, 0.001)\, GeV$ without
-with the lepton KK modes. It is observed that the BR
($\mu\rightarrow e \gamma$) is at the order of the magnitude of
$10^{-11}\, (10^{-12}, 10^{-13})$ for $\bar{\xi}^{E}_{N,\tau
e}=0.01\,(0.005, 0.001)\, GeV$ without the internal lepton KK mode
contributions. The addition of lepton KK modes bring considerable
enhancement, nearly $3\times$ one order (nearly two orders) for
set II (III), especially for the small values of the parameter
$k$. The larger enhancement in the case of set III is due to the
fact that the left handed leptons (zero and KK modes) are near to
the visible brane and their couplings to the new Higgs scalars,
living on the visible brane, become stronger. This is interesting
since it would be possible to enhance the BR by localizing the
lepton fields at appropriate points in the extra dimension, near
to the visible brane. Therefore, by choosing small values of
yukawa couplings, the experimental values of the BR of the LFV
decay under consideration could be reached.

\begin{figure}[htb]
\vskip -3.0truein \centering \epsfxsize=6.8in
\leavevmode\epsffile{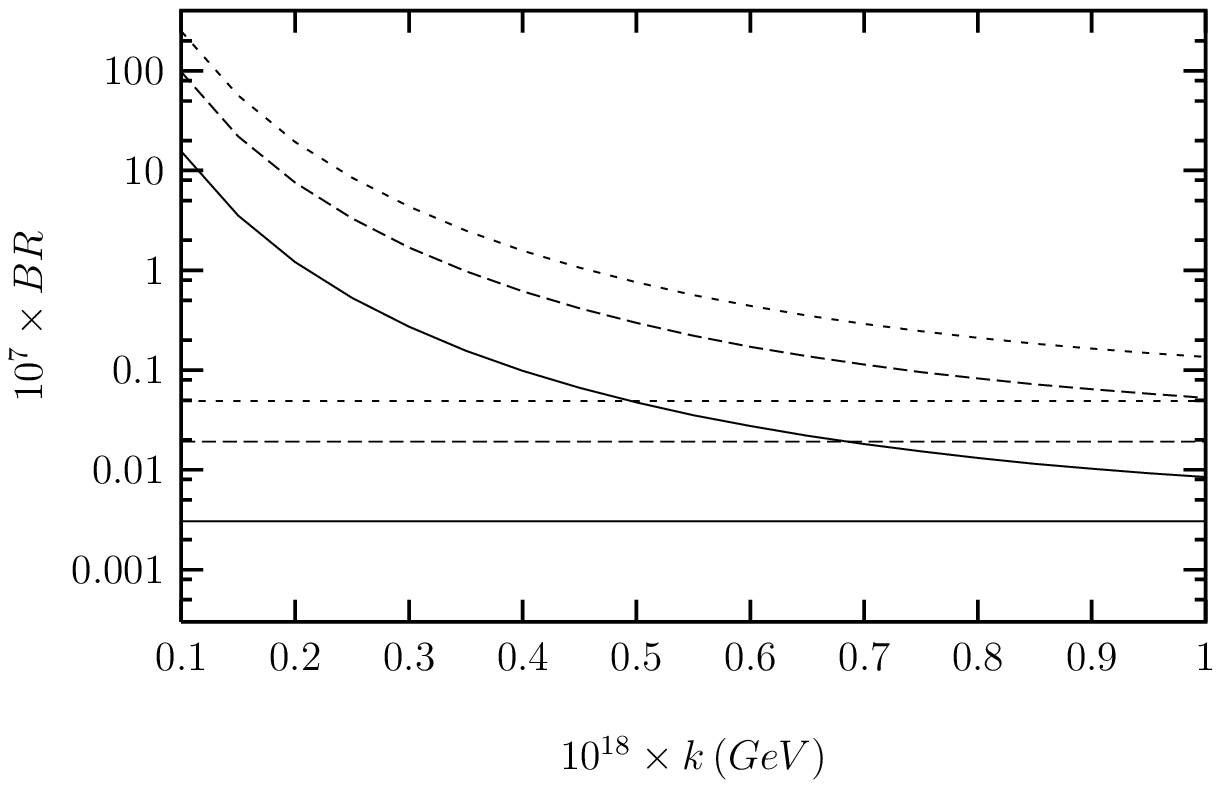} \vskip -3.0truein
\caption[]{ The parameter $k$ dependence of the BR of the  LFV
$\tau\rightarrow e \gamma$ decay for the location Set II, for
$\bar{\xi}^{E}_{N,\tau e}=1 \,GeV$ and for different values of the
coupling $\bar{\xi}^{E}_{N,\tau \tau}$. Here, the solid (dashed,
small dashed) line-curve represents the BR for and
$\bar{\xi}^{E}_{N,\tau \tau}=20\,(50, 80)\, GeV$ without-with the
lepton KK modes.} \label{BRtauegamkII}
\end{figure}

\begin{figure}[htb]
\vskip -3.0truein \centering \epsfxsize=6.8in
\leavevmode\epsffile{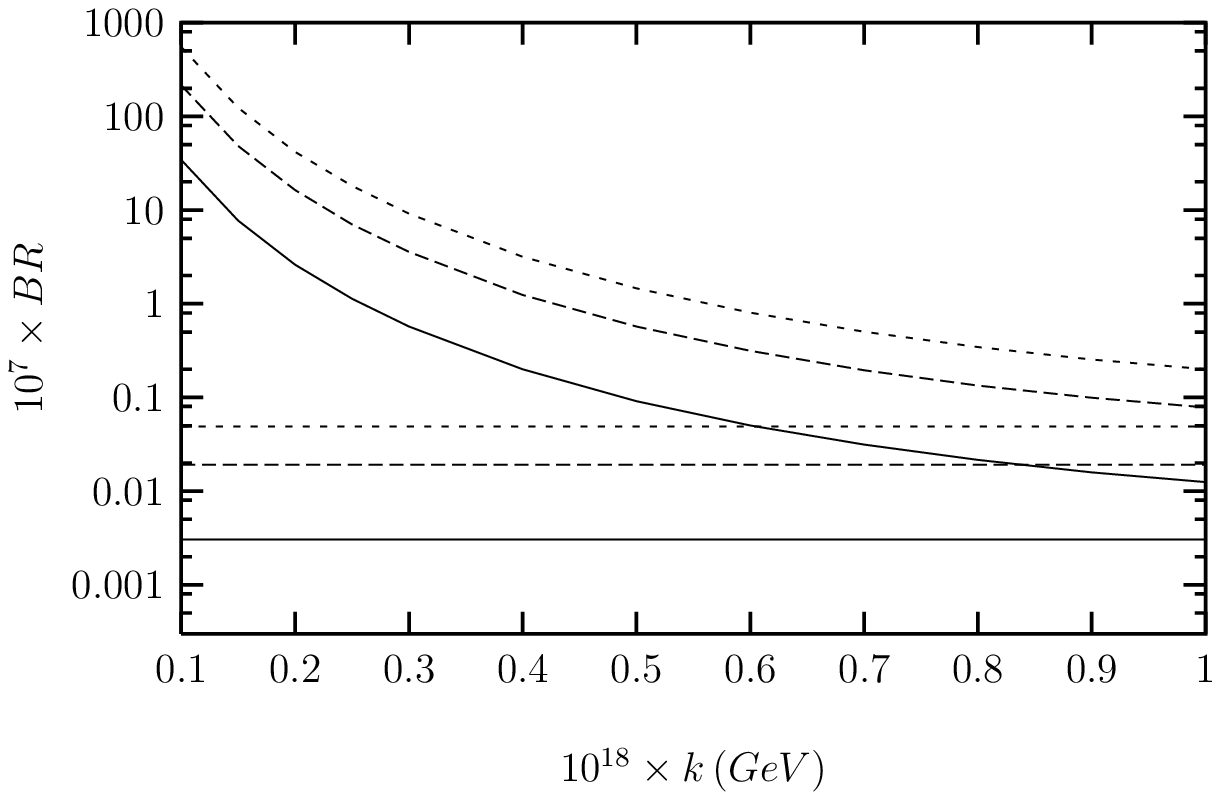} \vskip -3.0truein
\caption[]{ The same as Fig.\ref{BRtauegamkII} but for the
location Set III.} \label{BRtauegamkIId1}
\end{figure}

Figs. \ref{BRtauegamkII}-\ref{BRtauegamkIId1} represent the
parameter $k$ dependence of the BR of the LFV $\tau\rightarrow e
\gamma$ decay for lepton location set II-III, for
$\bar{\xi}^{E}_{N,\tau e}=1GeV$ and for different values of the
coupling $\bar{\xi}^{E}_{N,\tau \tau}$. Here, the solid (dashed,
small dashed) line-curve represents the BR for and
$\bar{\xi}^{E}_{N,\tau \tau}=20\,(50, 80)\, GeV$ without -with the
lepton KK modes. These figures show that the BR($\tau\rightarrow e
\gamma$) is at the order of the magnitude of $10^{-10}\, (10^{-9},
10^{-8})$ for $\bar{\xi}^{E}_{N,\tau \tau}=20\,(50, 80)\, GeV$
without the internal lepton KK mode contributions. With the
inclusion of the internal lepton KK modes, there occurs a
considerable enhancement, $5\times$ three order to almost $200\%$
(more than four orders to $300\%$) in the range of $k$, $0.1\leq k
\leq 1$ for set II (III). Even for the large curvature, the
enhancement of the BR due to the extra dimension contributions,
here, the internal lepton KK modes, is large and this is a worthy
observation for checking the existence and, also, the nature of
the extra dimensions.

\begin{figure}[htb]
\vskip -3.0truein \centering \epsfxsize=6.8in
\leavevmode\epsffile{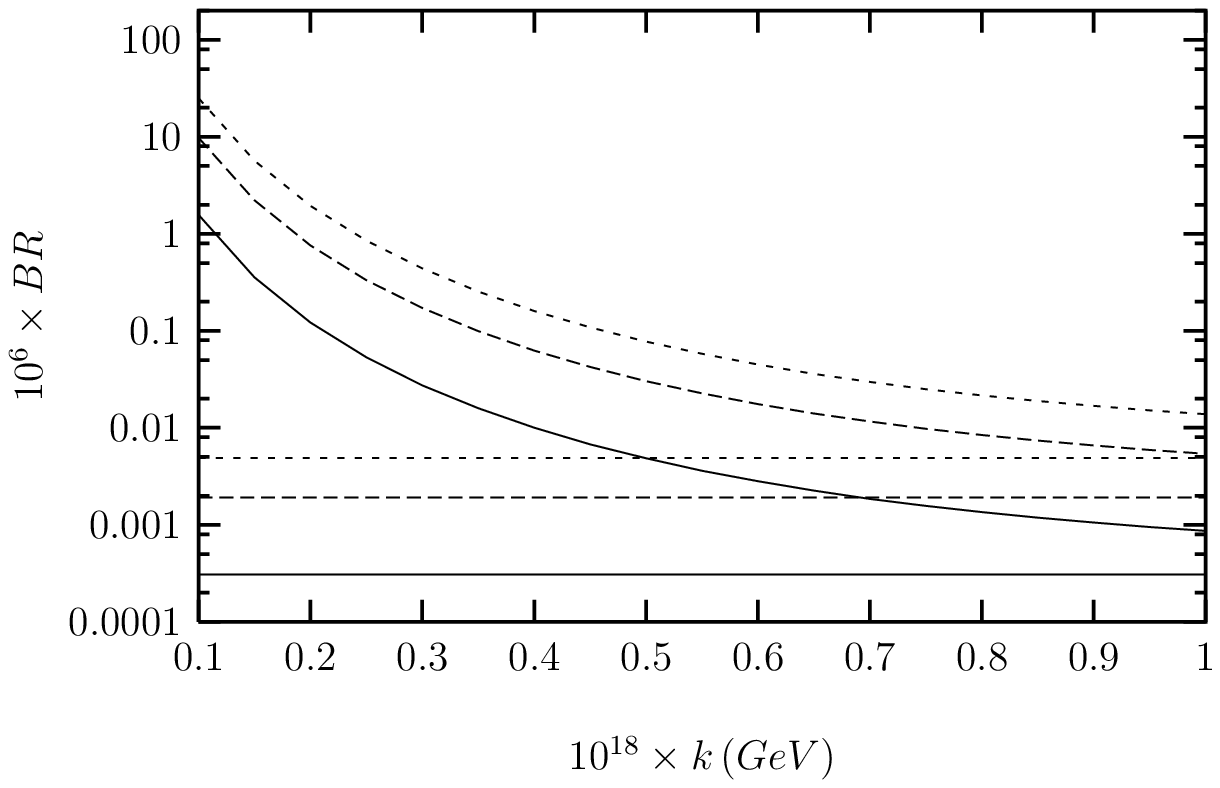} \vskip -3.0truein
\caption[]{ The parameter $k$ dependence of the BR of the  LFV
$\tau\rightarrow \mu \gamma$ decay for the location Set II, for
$\bar{\xi}^{E}_{N,\tau \mu}=1GeV$ and for different values of the
coupling $\bar{\xi}^{E}_{N,\tau \tau}$. Here, the solid (dashed,
small dashed) line-curve represents the BR for and
$\bar{\xi}^{E}_{N,\tau \tau}=20\,(50, 80)\, GeV$ without -with the
lepton KK modes.} \label{BRtaumugamkII}
\end{figure}

\begin{figure}[htb]
\vskip -3.0truein \centering \epsfxsize=6.8in
\leavevmode\epsffile{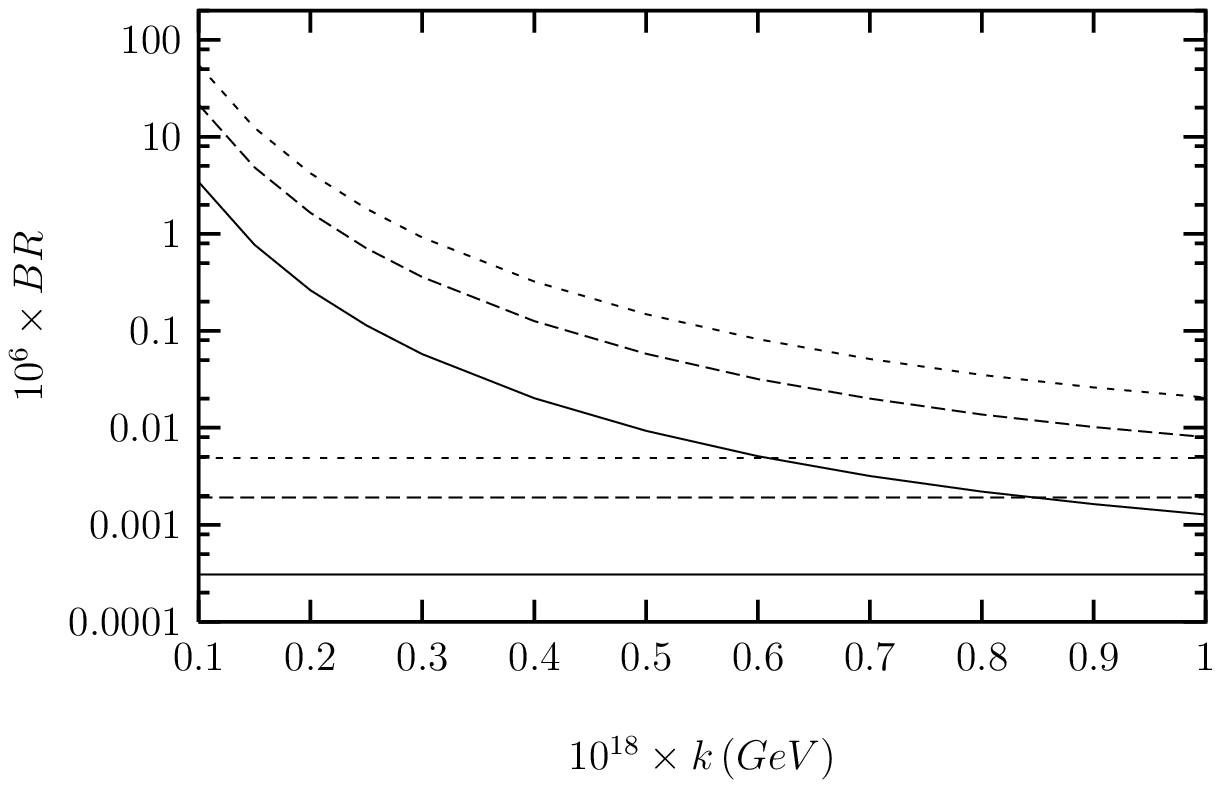} \vskip -3.0truein
\caption[]{ The same as Fig.\ref{BRtaumugamkII} but for the
location Set III.} \label{BRtaumugamkIId1}
\end{figure}

Finally, in Figs. \ref{BRtaumugamkII}-\ref{BRtaumugamkIId1}, we
present the parameter $k$ dependence of the BR of the LFV
$\tau\rightarrow \mu \gamma$ decay for lepton location set II-III,
for $\bar{\xi}^{E}_{N,\tau \mu}=1GeV$ and for different values of
the coupling $\bar{\xi}^{E}_{N,\tau \tau}$. Here, the solid
(dashed, small dashed) line-curve represents the BR for and
$\bar{\xi}^{E}_{N,\tau \tau}=20\,(50, 80)\, GeV$ without -with the
lepton KK modes. It is observed that the BR ($\tau\rightarrow \mu
\gamma$) is at the order of the magnitude of $10^{-10}\, (10^{-9},
10^{-8})$ for $\bar{\xi}^{E}_{N,\tau \tau}=20\,(50, 80)\, GeV$
without the internal lepton KK mode contributions. Similar to
previous decays, the inclusion of the internal lepton KK modes
bring a considerable enhancement to the BR, $5\times$ three orders
to almost $200\%$ (nearly more than four orders to more than
$300\%$) in the range of $k$, $0.1\leq k \leq 1$ for set II (III).
The current numerical result BR$=9.0\, (6.8)\, 10^{-8}$ at $90\%$
CL by \cite{Roney} (\cite{Aubert}) can be reached  even for weak
Yukawa interactions and for intermediate values of the curvature
parameter $k$, with the inclusion of lepton KK modes, especially
for the location set III.

For completeness, we present the results of the same processes for
various scenarios in the case of flat extra dimensions. For the
flat one universal extra dimension, the enhancements of the BRs of
the LFV decays (especially the BR of the $\mu\rightarrow e \gamma$
decay), due to the lepton KK modes and the new neutral Higgs KK
modes are tiny for the compactification radius $R$, $1/R\geq 500\,
GeV$. However, since the contribution of the universal extra
dimension to any process exists at least in the loop
level\footnote{The processes with incoming and outgoing zero mode
particles can get contributions from the universal extra dimension
at least in the loop level since the KK number should be conserved
at the vertices and the tree level contribution with a single KK
mode does not appear.}, the LFV decays, which can exist also in
the loop level and which are clean in the sense that are free from
the strong interactions, are among the most informative processes
in order to search this type of extra dimension. If one considers
the flat non-universal extra dimensions where the new Higgs bosons
are accessible to the extra dimension but leptons live on the 4D
brane\footnote{The electric dipole moments of the charged leptons
are among the clean physical quantities and they are worthwhile to
study in order to search the extra dimensions.}, there exist
considerable enhancements in the BRs of the LFV decays, in the
case of two spatial extra dimensions and the discrepancy between
the experimental and the theoretical values of the BRs can be
overcome \cite{IltanExtrDim, IltanExtrDim1}. In the the split
fermion scenario where the fermions are localized with Gaussian
profiles at different points in the extra dimension, the
enhancements in the BRs of LFV decays are tiny for a single extra
dimension and the sensitivities of the BRs increase for two extra
dimensions \cite{IltanLFVSplit}. For both cases, the enhancements
in the BRs are due to the abundance of new neutral Higgs KK modes.
If  the new Higgs doublet is restricted to the 4D brane (thin
bulk), we observe that the BRs are sensitive to the location of
the 4D brane in the extra dimension (the width of the thin bulk),
especially for the $\mu\rightarrow e \gamma$ decay
\cite{IltanLFVFatSplit}. Notice that in the case of non-universal
extra dimensions, there is no need for the KK number to conserve
at the vertices and the processes with a single internal KK mode
can exist. Therefore, the non-universal extra dimension(s) results
in non zero contribution also in the tree level and the physical
quantities related to such processes are not suppressed. Even if
the BRs of the LFV interactions are sensitive to the effects of
the non-universal extra dimensions, the processes existing in the
tree level with internal KK modes are more informative to search
the extra dimensions.

Now we would like to present our results briefly.
\begin{itemize}

\item The idea that the hierarchy of new Yukawa couplings is
described by the location of leptons in the bulk of RS1 background
is worthwhile to study. However, with a unique set of Yukawa
couplings, the coupling $\alpha$ and the others created from
$\alpha$,  we observe that it is not possible to get theoretical
values of the BRs coinciding with the current experimental results
of the LFV decays under consideration. Furthermore, we see that
the BRs are not sensitive to the curvature parameter $k$ and the
internal lepton KK mode contributions. Notice that, in this case,
we choose the left handed and right handed charged lepton
positions as the same in the extra dimension. The future more
accurate experimental measurements, especially for the
$\tau\rightarrow e \gamma$ decay, would make it possible to test
the physics behind the hierarchy of new Yukawa couplings and the
effect of extra dimensions on this hierarchy.
\item In the case that the left handed and right handed charged
lepton positions are different in the extra dimension, the
addition of lepton KK modes result in considerable enhancement in
the BRs, especially for the small values of the parameter $k$.
Furthermore, this enhancement increases and the sensitivity of BR
to the curvature parameter $k$ becomes strong by taking the
location points of left handed charged leptons near to the visible
brane. This is due to the fact that the left handed KK mode
couplings to the new Higgs scalars, living on the visible brane,
become stronger. Therefore, it would be possible to enhance the
BRs by localizing the lepton fields at appropriate points in the
extra dimension, and, even with the small values of Yukawa
couplings and even in  single extra dimension, the experimental
values of the BR of the LFV decay under consideration can be
reached. This is interesting and the forthcoming measurements of
the BRs of the LFV decays, especially the $\mu\rightarrow e
\gamma$ decay, would give considerable information about the
existence and, also, the nature of the warped extra dimensions. .
\end{itemize}
\newpage
\newpage
\end{document}